\documentclass[floatfix, preprint, showpacs, showkeys, preprintnumbers, nofootinbib, superscriptaddress]{revtex4-1}
\usepackage{comment}
\usepackage{amssymb}
\usepackage{color}
\usepackage{rotating}
\usepackage{amsmath}
\usepackage{ulem}
\usepackage{overpic}
\usepackage{graphicx}
\usepackage{dcolumn}
\usepackage{bm}
\usepackage{chngpage}
\usepackage{multirow}
\usepackage{slashed}
\usepackage{indentfirst}
\usepackage{booktabs}
\usepackage{amssymb,amsfonts}
\usepackage{amsmath}
\usepackage{color}
\usepackage[colorlinks,citecolor=blue,anchorcolor=blue,linkcolor=blue,hypertex,breaklinks=true]{hyperref}

\begin{document}
\title{Evolution of nuclear charge radii in copper and indium isotopes}
\author{Rong An}
\affiliation{Key Laboratory of Beam Technology of Ministry of Education, Institute of Radiation Technology, Beijing Academy of Science and Technology, Beijing 100875, China}
\affiliation{Key Laboratory of Beam Technology of Ministry of Education,
College of Nuclear Science and Technology, Beijing Normal University, Beijing 100875, China}

\author{Xiang Jiang}
\affiliation{College of Physics and Optoelectronic Engineering, Shenzhen University, Shenzhen 518060, China}

\author{Li-Gang Cao}
\affiliation{Key Laboratory of Beam Technology of Ministry of Education,
College of Nuclear Science and Technology, Beijing Normal University, Beijing 100875, China}
\affiliation{Key Laboratory of Beam Technology of Ministry of Education, Institute of Radiation Technology, Beijing Academy of Science and Technology, Beijing 100875, China}

\author{Feng-Shou Zhang}
\email[E-mail: ]{fszhang@bnu.edu.cn}
\affiliation{Key Laboratory of Beam Technology of Ministry of Education, Institute of Radiation Technology, Beijing Academy of Science and Technology, Beijing 100875, China}
\affiliation{Key Laboratory of Beam Technology of Ministry of Education,
College of Nuclear Science and Technology, Beijing Normal University, Beijing 100875, China}
\affiliation{Center of Theoretical Nuclear Physics,
National Laboratory of Heavy Ion Accelerator of Lanzhou, Lanzhou 730000, China}


\begin{abstract}
  Systematic trends in nuclear charge radii are of great interest due to universal shell effects and odd-even staggering (OES). The modified root mean square (rms) charge radius formula, which phenomenologically accounts for the formation of neutron-proton ($np$) correlations, is here applied for the first time to the study of odd-$Z$ copper and indium isotopes. Theoretical results obtained by the relativistic mean field (RMF) model with NL3, PK1 and NL3$^{*}$ parameter sets are compared with experimental data. Our results show that both OES and the abrupt changes across $N=50$ and $82$ shell closures are clearly reproduced in nuclear charge radii. The inverted parabolic-like behaviors of rms charge radii can also be described remarkably well between two neutron magic numbers, namely $N=28$ to $50$ for copper isotopes and $N=50$ to $82$ for indium isotopes. This implies that the $np$-correlations play an indispensable role in quantitatively determining the fine structures of nuclear charge radii along odd-$Z$ isotopic chains. Also, our conclusions have almost no dependence on the effective forces.
\end{abstract}


\maketitle
\section{INTRODUCTION}\label{sec0}
The nuclear charge radius $R_\mathrm{ch}$ is one of the most indispensable parameters in characterizing the fundamental properties of finite nuclei over the entire periodic table. The evolution of nuclear charge radii along isotopic chains exhibits particular features, such as odd-even staggering (OES) and abrupt changes across neutron shell closures~\cite{Angeli_2009,ANGELI201369,GarciaRuiz:2019cog,LI2021101440}. Knowledge of these local variations gives more insight into the information about charge density distributions $\rho_{\mathrm{ch}}(\mathbf{r})$ and Coulomb interactions in nuclei. Especially, various nuclear structure phenomena are associated with charge radii, such as halo structures, shape coexistence and onset of deformation~\cite{PhysRevLett.102.062503,RPL2013212501,PhysRevC.90.014317,SELIVERSTOV2013362}. Also, direct determination of the neutron-skin thickness, which is known as one of the most sensitive terrestrial probes of the symmetry energy in the equation of state, usually involves the precise measurement of the charge density distributions~\cite{PhysRevLett.85.5296,PhysRevLett.86.5647,PhysRevLett.95.122501,PhysRevC.72.064309,PhysRevLett.102.122502,PhysRevLett.108.112502,Carbone2010,PhysRevC.82.024321,
PhysRevC.85.032501,ZHANG2013234,PhysRevC.102.044316,CAOLG2015,Roca2015,Behera_2020,GAIDAROV2020122061,PhysRevLett.126.172502}. Thus, the accurate description of nuclei charge radii plays an essential role in experimental and theoretical studies.

Plenty of methods are employed to derive the nuclear size, such as high-energy elastic electron scattering ($e^{-}$)~\cite{PhysRev.92.978,PhysRevC.21.1426}, muonic atom X-rays ($u^{-}$)~\cite{Engfer:1973df,FRICKE1995177,BAZZI2011199}, high-resolution laser spectroscopy~\cite{PhysRevLett.93.113002}, highly-sensitive Collinear Resonance Ionization Spectroscopy (CRIS)~\cite{Cocolios:2013wpa,vernon2020laser}, optical isotope shifts (OIS) and K$_{\alpha}$ X-ray isotope shifts (K$_{\alpha}$IS)~\cite{ANGELI2004185}. Furthermore, the values of charge radii can also be extracted indirectly from the experimental observables of $\alpha$-decay properties~\cite{Dongdong2013,Qian2014,QIAN2016134,Qian2018,Manjunatha2020} and cluster or proton emission data~\cite{Qian2013}. With the accumulation of detected data far away from the $\beta$-stability line, considerable efforts have been undertaken in recent years to
describe the observed regular and universal behaviors of nuclear charge radii along isotopic chains.

In general, the charge radius is ruled by the $A^{1/3}$ law along the stability line~\cite{A.Bohr,P.Ring}. However, pronounced discrepancies between experimental data and theoretical results have been gradually found toward neutron- or proton-rich regions~\cite{ANGELI201369}. An improved empirical formula including isospin and shell corrections has been proposed, with which the root mean square (rms) deviation $\sigma$ in nuclear charge radii falls to 0.022 fm~\cite{PhysRevC.88.011301}. Also, an effective five-parameter formula considering the Casten factor and OES effect provides an almost comparable rms deviation, $\sigma=0.0223$ fm~\cite{Sheng:2015poa}. Another set of phenomenological formulas, called Garvey-Kelson (GK) relations, produce an rms deviation $\sigma\approx0.01$ fm~\cite{Piekarewicz:2009av}. However, the GK relations rely on the adjacent nuclei of known size~\cite{sun2014,zhao2016}, which results in limited extrapolative ability~\cite{liu2011,cheng2014}. Another alternative form as a function of $Z^{1/3}$ has also been proposed, with which the rms deviation decreases to $0.007$ fm~\cite{Zhang:2001nt}, and corresponding improved versions have also been derived~\cite{KNWL200203009,Lei2007}. A recent study shows that the considered $\delta{R}_{i{\rm{n}}-j{\rm{p}}}$ relations ($i, j =1$ or $2$) also produce a similar rms deviation ($\sigma=0.0072$ fm) in nuclear charge radii~\cite{Bao2020PRC}. 

The mean-field approach, such as the non-relativistic Hartree-Fock-Bogoliubov (HFB) model~\cite{Goriely2009,Goriely2010} and relativistic mean-field (RMF) theory~\cite{Geng:2003pk,zhao2010}, can describe nuclear charge radii in a self-consistent way. Both kinds of models can produce the nuclear binding energies well, but quantitatively fail to follow the fine structure of charge radii along isotopic chains. $Ab~initio$ many-body calculations with chiral effective field theory (EFT) interactions encounter the same issues~\cite{BINDER2014119,PhysRevC.91.051301,Ruiz2016,PhysRevC.96.014303}. Recently developed Bayesian neural networks can produce powerful predictions for nuclear charge radii~\cite{Utama_2016,RenZZ2020,Wu:2020bao}. However, the detailed trend of local variations, especially OES behaviors, cannot be reproduced well. In Ref.~\cite{Dong:2021aqg}, the OES in charge radii has been supplemented into a Bayesian neural network. The strong OES behaviors of nuclear charge radii are modelled very well for calcium isotopes.

Many possible mechanisms are employed to investigate the systematic evolution of nuclear charge radii. The intriguing features of OES and inverted parabolic-like behaviors in nuclear charge radii are clearly observed along the calcium isotopic chain~\cite{ANGELI201369,Ruiz2016,Miller2019}. A good fit result for neutron-proton radii differences can be obtained if the core polarization caused by valence neutrons is taken into account theoretically~\cite{CAURIER198015,TALMI1984189}. For regions beyond $N=29$, the results are not so clear. In Ref.~\cite{Reinhard2017}, the sophisticated Fayans EDF model which considers a novel density-gradient term in pairing interactions can almost reproduce the OES of charge radii in calcium isotopes. This means surface pairing interactions play a dominant role in reproducing the fine structure of nuclear charge radii. Nevertheless, a large deviation still exists beyond $N>30$. The latest study on mercury isotopes shows that nuclear charge radii can be described predominately at the mean-field level, in which pairing does not need to play a crucial role in the origins of OES~\cite{Goodacre2021}. This may suggest that many-body correlations can be gradually captured by a mean-field approach with increasing mass number.

Other possible mechanisms may have an influence on nuclei size, such as quadrupole vibrations~\cite{REEHAL1971385}, $\alpha$-particle cluster~\cite{Zawischa:1985qds}, three- or four-body interactions~\cite{ZAWISCHA1987299,PhysRevLett.61.149}, higher order radial moments $\langle{r^{4}}\rangle$~\cite{Reinhard2020PRC,reinhard2021nuclear},  deformation~\cite{GIROD19821,Ulm:1986wd,An:2021rlw}, and especially the isospin symmetry breaking results in the violation of Coulomb interactions~\cite{Lam2013PRC}. As argued in Ref.~\cite{Miller:2018mfb}, the influence of isospin symmetry breaking needs to be considered carefully in theoretical studies. This reflects that the neutron-proton ($np$) correlations at low density regions make contributions to the nuclear charge radii. In fact, the formation of the OES in nuclear charge radii due to four-particle correlations suggests that $np$-correlations play an indispensable role~\cite{Zawischa:1985qds}. A recent study has determined that the role of $np$-correlations have a non-negligible impact on nuclear size~\cite{Ryckebusch2021}.

As mentioned above, changes to nuclear size are determined by many possible mechanisms. Therefore, a unified approach is necessary to describe the systematic evolution of nuclear charge radii. In contrast to Fayans EDF model, a phenomenologically-modified charge radius formula associated with Cooper pair condensation has been established in RMF theory~\cite{An:2020qgp}. This proposed model, namely the RMF(BCS)$^{*}$ approach, yields results consistent with experimental data not only for calcium isotopes, but also for $10$ other even-$Z$ isotopic chains. It is necessary to further check our ansatz along odd-$Z$ isotopic chains. In this work, charge radii of copper and indium isotopes are studied. Precise knowledge of nuclear charge radii is essential for the understanding of the nuclear force~\cite{reinhard2021nuclear}. Therefore, theoretical results obtained by NL3, PK1 and NL3$^{*}$ parameter sets are compared with experimental data. The results obtained from a non-relativistic Skyrme Hartree-Fock model with the SV-bas force are also shown for comparison~\cite{MARUHN20142195}.


This paper is organized as follows. In Sec.~\ref{sec1}, a brief outline of the theory is given. In Sec.~\ref{sec2}, the numerical results and discussions are provided. Finally, a summary is given in Sec.~\ref{sec3}.

\section{THEORETICAL FRAMEWORK}\label{sec1}
Relativistic mean field (RMF) theory with different parameter sets has made considerable success in describing ground or low-excited state properties of finite nuclei~\cite{MENG19983,Vretenar:2005zz,Liang:2014dma,jie2016relativistic,PhysRevC.67.034312,PhysRevC.69.054303,zhang2007,PhysRevC.90.044305,zhang2012,PhysRevC.92.024324,Cao:2003yn,An:2020wcv,PhysRevC.102.024314,
Wang:2021bht,PhysRevC.68.034323,PhysRevC.82.011301,SUN2018530,PhysRevC.104.L021301,Wang:2021tjg}. In the nonlinear self-consistent Lagrangian density, nucleons are described as Dirac particles which interact via the exchange of $\sigma$, $\omega$ and $\rho$ mesons. The electromagnetic field is mediated by photons. The effective Lagrangian density is defined as follows,
\begin{eqnarray}\label{lag1}
\mathcal{L}&=&\bar{\psi}[i\gamma^\mu\partial_\mu-M-g_\sigma\sigma
-\gamma^\mu(g_\omega\omega_\mu+g_\rho\vec
{\tau}\cdotp\vec{\rho}_{\mu}+e\frac{1-\tau_{3}}{2}A_\mu)]\psi\nonumber\\
&&+\frac{1}{2}\partial^\mu\sigma\partial_\mu\sigma-\frac{1}{2}m_\sigma^2\sigma^2
-\frac{1}{3}g_{2}\sigma^{3}-\frac{1}{4}g_{3}\sigma^{4}\nonumber\\
&&-\frac{1}{4}\Omega^{\mu\nu}\Omega_{\mu\nu}+\frac{1}{2}m_{\omega}^2\omega_\mu\omega^\mu+\frac{1}{4}c_{3}(\omega^{\mu}\omega_{\mu})^{2}\nonumber\\
&&-\frac{1}{4}\vec{R}_{\mu\nu}\cdotp\vec{R}^{\mu\nu}+\frac{1}{2}m_\rho^2\vec{\rho}^\mu\cdotp\vec{\rho}_\mu
+\frac{1}{4}d_{3}(\vec{\rho}^{\mu}\vec{\rho}_{\mu})^{2}\nonumber\\
&&-\frac{1}{4}F^{\mu\nu}F_{\mu\nu},
\end{eqnarray}
where $M$ is the mass of nucleon and $m_{\sigma}$, $m_{\omega}$, and $m_{\rho}$, are the masses of the $\sigma$, $\omega$ and $\rho$ mesons, respectively. Here, $g_{\sigma}$, $g_{\omega}$, $g_{\rho}$, $g_{2}$, $g_{3}$, $c_{3}$ and $d_{3}$ are the coupling constants for $\sigma$, $\omega$ and $\rho$ mesons.
The field tensors for the vector mesons and photon fields are defined as $\Omega_{\mu\nu}=\partial_{\mu}\omega_{\nu}-\partial_{\nu}\omega_{\mu}$,
$\vec{R}_{\mu\nu}=\partial_{\mu}\vec{\rho}_{\nu}-\partial_{\nu}\vec{\rho}_{\mu}-g_{\rho}(\vec{\rho}_{\mu}\times\vec{\rho}_{\nu})$ and $F_{\mu\nu}=\partial_{\mu}A_{\nu}-\partial_{\nu}A_{\mu}$.

The Dirac equations can be deduced from Eq.~(\ref{lag1}) and expressed as
\begin{eqnarray}
\{-i\alpha\nabla+V(\mathbf{r})+\beta{M^{*}}\}\psi_{i}=\varepsilon_{i}\psi_{i},
\end{eqnarray}
where $M^{*}=M+S(\mathbf{r})$ is the effective mass of the nucleon, which is significantly smaller and space dependent because of the $\sigma$-meson field~\cite{PhysRevC.44.2552}. The fields $S(\mathbf{r})$ and $V(\mathbf{r})$ are connected in a self-consistent way to densities. $\psi_{i}$ represents the spinor wave function, which yields the expectation values of total energies, root-mean-square (rms)  radii, etc. The corresponding Dirac equations for nucleons and Klein-Gordon equations for mesons and photons are solved by the expansion method with the harmonic oscillator basis~\cite{Geng:2003pk,Ring:1997tc}. In the present work, 12 shells are used for expanding the fermion fields and 20 shells for the meson fields.

In the conventional RMF model, the mean-square charge radius is calculated in the following way (in units of fm$^2$)~\cite{Ring:1997tc,Geng:2003pk}:
\begin{eqnarray}\label{coop1}
R_{\mathrm{ch}}^{2}=\frac{\int{r}^{2}\rho_{\mathrm{p}}(\mathbf{r})d^{3}r}{\int\rho_{\mathrm{p}}(\mathbf{r})d^{3}r}+0.64~\mathrm{fm^{2}},
\end{eqnarray}
where the first term represents the charge density distribution of point-like protons and the last term takes into account the finite size effect of the proton~\cite{Ring:1997tc}. To account for the experimentally-observed odd-even staggering of nuclear charge radii, a modified formula has been proposed in Ref.~\cite{An:2020qgp}, which is written as (in units of fm$^2$)
\begin{eqnarray}\label{cp1}
R_{\mathrm{ch}}^{2}=\frac{\int{r}^{2}\rho_{\mathrm{p}}(\mathbf{r})d^{3}r}{\int\rho_{\mathrm{p}}(\mathbf{r})d^{3}r}+0.64~\mathrm{fm^{2}}+\frac{a_{0}}{\sqrt{A}}\Delta{\mathcal{D}}~\mathrm{fm^{2}}.
\end{eqnarray}
In this expression, the last term on the right hand is related to the Cooper pairs condensation~\cite{PhysRevC.76.011302}. The quantity $A$ is the mass number and $a_{0}$ is a normalization constant. The quantity $\Delta\mathcal{D}=|\mathcal{D}_{\mathrm{n}}-\mathcal{D}_{\mathrm{p}}|$ represents the difference in Cooper pairs condensation between neutrons and protons. It is calculated self-consistently by solving the state-dependent BCS equations~\cite{Geng:2003pk,RongAn:114101}.
More details are shown in Ref.~\cite{An:2020qgp}.

\section{RESULTS AND DISCUSSIONS}\label{sec2}
\subsection{Shell effects of charge radii for Cu and In isotopes}
In this work, the nuclear charge radii along copper and indium isotopic chains are systematically investigated. The pairing strength is generally determined through the empirical odd-even mass staggering~\cite{Bender:2000xk}. The pairing strength $V_{0}=350$ MeV fm$^{3}$ is employed for the NL3 set, but $V_{0}=380$ MeV fm$^{3}$ is used for the PK1 and NL3$^{*}$ forces. The parameter $a_{0}$ involved in Eq.~(\ref{cp1}) is hereto consistent with Ref.~\cite{An:2020qgp}, $0.834$ for copper isotopes and $0.22$ for indium isotopes. To facilitate the following discussion, the results obtained by Eq.~(\ref{coop1}) are denoted as RMF(BCS), while RMF(BCS)$^{*}$ represents the modified results derived from Eq.~(\ref{cp1}).

\begin{figure}[htbp]
\includegraphics[scale=0.45]{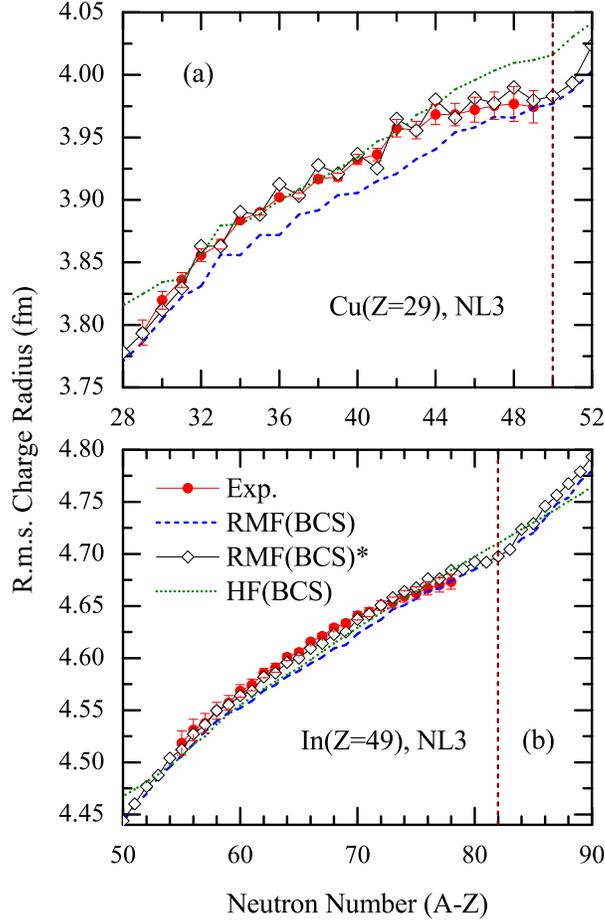}
\caption{(color online) Charge radii of copper (a) and indium (b) isotopes with (open diamonds) and without (dashed line) modified formula are plotted in the relativistic mean field (RMF) framework with the NL3 parameter set. The experimental data (solid circles) stem from Refs.~\cite{ANGELI201369,deGroote,Sahoo_2020,LI2021101440}. The results obtained by Hartree-Fock plus BCS approach with SV-bas Skyrme parametrization, labelled HF(BCS) (dotted line), are also shown for comparison~\cite{MARUHN20142195}.} \label{figch}
\end{figure}
In Fig.\ref{figch}, the rms charge radii of copper and indium isotopes are obtained by the RMF(BCS) and RMF(BCS)$^{*}$ approaches. It is clearly shown that the RMF(BCS) method is unable to give a satisfactory trend of charge radii along copper isotopes, especially the rising trend (defined as inverted parabolic-like behavior~\cite{perera2021charge}) in the region between $N=28$ and $N=50$. Considering the modified term, the RMF(BCS)$^{*}$ approach reproduces the experimental data well. However, the odd-even oscillation behavior is slightly overestimated in $^{65-73}$Cu isotopes, which may be attributed to the last unpaired proton and neutron that violate the time-reversal symmetry~\cite{P.Ring}. In our calculations, the neutron-proton ($np$) correlations coming from unpaired neutrons and protons are not taken into account. The same scenario can also be found in the odd-$Z$ potassium isotopes where the OES is still overestimated in charge radii~\cite{An:2021rlw}. As a comparison, the results obtained by the HF(BCS) approach with SV-bas Skyrme parametrization are also shown. The HF(BCS) model seems to reproduce the tiny variations between $N=31$ and $N=44$ region, but larger deviation occurs in neutron (proton)-rich regions, particularly around $N=28$ and $N=50$. The shell effects can be described well due to the rather small isospin dependence of the spin-orbit interactions~\cite{Sharma1995PRL}. This means the isospin dependence of the spin-orbit interactions are overestimated in the Skyrme Hartree-Fock model for the SV-bas force.

As shown in Fig.~\ref{figch}(b), charge radii of indium isotopes obtained by the RMF(BCS) approach slightly deviate from the experimental data between $A=104$ and $A=122$. These results are similar to those obtained by HF(BCS) model. As encountered in copper isotopes, HF(BCS) still overestimates the rms charge radii across neutron shell closures. The inverted parabolic-like behavior can also be reproduced well by the RMF(BCS)$^{*}$ approach. As is well known, this inverted parabolic-like behaviors of charge radii between two strong filled-shells are observed generally along isotopic chains, such as in calcium~\cite{Ruiz2016}, cadmium~\cite{Hammen2018}, tin~\cite{Gorges2019}, etc. Ref.~\cite{Sun2017PRC} points out that the parabolic-like behavior of charge radii is formed due to a linear correlation between charge radius and the corresponding quadrupole deformation. In our study, the $np$-correlations associated with Cooper pairs condensation make a dominant contribution.

These nuclei with $N=50$ and $N=82$ closed-shells are more difficult to be excited than their neighbors, which is evidenced by their relatively high excitation energies and low excitation probabilities~\cite{sunshuai2021,CORTES2020135071,Steppenbeck2013}. The other intriguing phenomenon is a sudden increase in the slope of the change of charge radius across the magic numbers~\cite{ANGELI201369,GarciaRuiz:2019cog}, known as the kink. The abrupt change of charge radii beyond the $N=50$ neutron shell closure is shown in Fig.~\ref{figch}(a). A similar case was also studied earlier in RMF within the NL-SH parametrization set~\cite{LALAZISSIS1995201}. In Fig.~\ref{figch}(b), this rapid increase can still be reproduced by the RMF(BCS)$^{*}$ approach across the magic number $N=82$. These characteristic phenomena are found dramatically in the latest studies of cadmium~\cite{Hammen2018}, tin~\cite{Gorges2019}, mercury~\cite{Goodacre2021}, etc. The shell closures have been identified by this discontinuity in the slope of the change of rms charge radii as a function of the neutron numbers along isotopic chains. As mentioned in Ref.~\cite{Gorges2019}, the pronounced kink at the $N = 82$ shell results from the reduced neutron pairing. This has been attributed to the rather small isospin dependence of the spin-orbit term in the RMF model~\cite{Sharma1995PRL}.
\begin{figure}[htbp]
\includegraphics[scale=0.6]{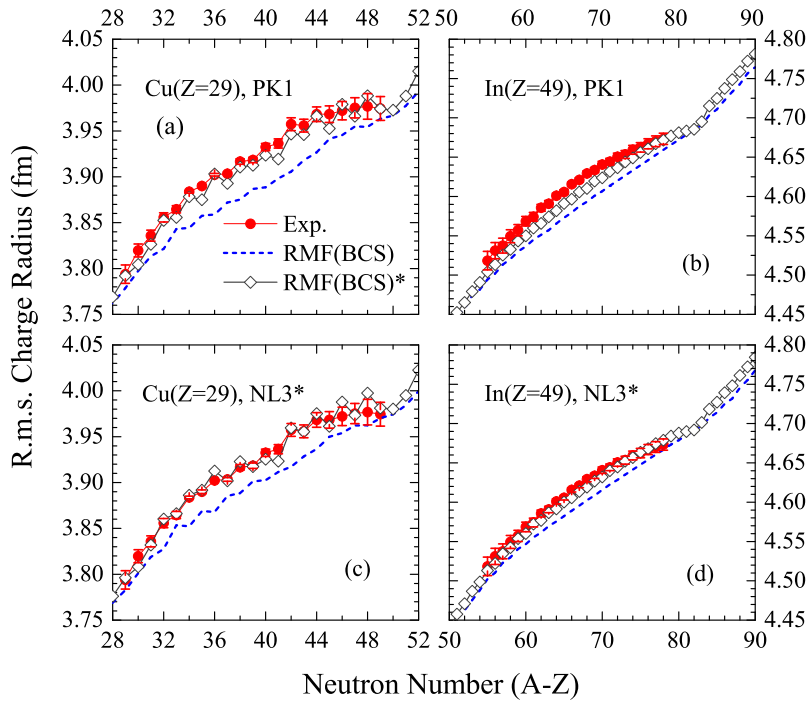}
\caption{(color online) Charge radii of copper (a, c) and indium (b, d) isotopes obtained by effective forces PK1~\cite{PhysRevC.69.034319} and NL3$^{*}$~\cite{LALAZISSIS200936}. The labels are the same as for Fig~\ref{figch}. The experimental data stem from Refs.~\cite{ANGELI201369,deGroote,Sahoo_2020,LI2021101440}. } \label{figch11}
\end{figure}

Precise knowledge of nuclear charge radii plays an critical role in understanding the nuclear force~\cite{reinhard2021nuclear}. In Fig.~\ref{figch11}, therefore, charge radii of copper and indium isotopes obtained by PK1~\cite{PhysRevC.69.034319} and NL3$^{*}$~\cite{LALAZISSIS200936} forces are presented. From the figure, one can find both parameter sets describe the inverted parabolic-like behaviors well for copper and indium isotopes if the $np$-correlations are taken into account. Moreover, the abrupt changes of charge radii are matched remarkably well across the $N=50$ and $N=82$ neutron magic shells. However, the odd-even oscillations of nuclear charge radii are overestimated in copper isotopes. Also, the modified results with the PK1 force slightly diverge with respect to the experimental data for $^{109-119}$In isotopes. Overall, the rms charge radii of nuclei can be described well by the NL3, PK1 and NL3$^{*}$ parameter sets only if the contributions of $np$-correlations are considered. The independence of the results from the effective forces means that $np$-correlations are the essential ingredient in describing the rms charge radii. This is in agreement with the conclusion of Ref.~\cite{An:2020qgp}.

\subsection{OES in nuclear charge radii for Cu and In isotopes}
\begin{figure}[htbp]
\includegraphics[scale=0.45]{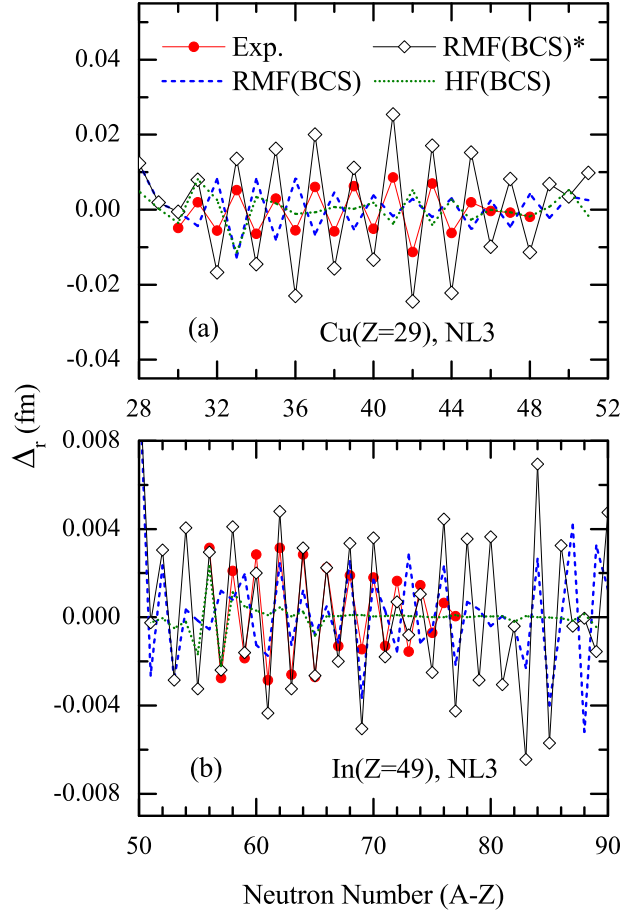}
    \caption{(color online) Odd-even staggering of nuclear charge radii obtained by the RMF(BCS) (dashed line) and RMF(BCS)$^{*}$ (open diamonds) approaches with the NL3 force along copper (a) and indium (b) isotopic chains. The experimental data (solid circles) stem from Refs.~\cite{ANGELI201369,deGroote,Sahoo_2020,LI2021101440}. The results obtained by Hartree-Fock plus BCS approach with SV-bas Skyrme parametrization, labelled HF(BCS) (dotted line), are also shown for comparison~\cite{MARUHN20142195}.} \label{figoe}
\end{figure}
As is well known, OES of nuclear masses is generally observed throughout the nuclear chart~\cite{Wangmeng30002}. The same situation is encountered in nuclear charge radii~\cite{Angeli_2009,ANGELI201369,GarciaRuiz:2019cog}. This staggering describes the fact that the nuclear charge radii of odd-$N$ isotopes are smaller than the averages of their even-$N$ neighbours. In order to emphasize these local variations, a three-point formula has been employed to extract the odd-even oscillation behaviors of nuclear charge radii along a specific isotopic chain~\cite{Reinhard2017}. The definition is written as follows,
\begin{eqnarray}\label{oef}
\Delta_{r}(N,Z)=\frac{1}{2}[R(N-1,Z)-2R(N,Z)+R(N+1,Z)],
\end{eqnarray}
where $R(N,Z)$ is rms charge radius for a nucleus with neutron number $N$ and proton number $Z$.

In Fig.~\ref{figoe}, the degree of OES of the nuclear charge radii along the copper and indium isotopic chains is plotted as a function of neutron numbers. The results obtained by the HF(BCS) method with SV-bas Skyrme parametrization are also shown for comparison~\cite{MARUHN20142195}. In this figure, the RMF(BCS)$^{*}$ approach is able to reproduce the OES of nuclear charge radii along these two isotopic chains. However, this modified expression slightly overestimates the results along the copper isotopic chain. The trend of comparison is also evidently shown in Fig.~\ref{figch}(a) where the amplitudes of OES in nuclear charge radii are slightly exaggerated. The same scenario is reported in Ref.~\cite{An:2021rlw}, where the OES is also overestimated along the potassium isotopic chain. We deem that this discrepancy is associated with unpaired nucleons~\cite{Ulm:1986wd,Lievens:1995iz,AHMAD1988244}. In the present study, the $np$-correlations coming from the unpaired neutron and proton are excluded. One can also find that both RMF(BCS) and HF(BCS) models cannot reproduce the OES well, and even give inverted trends. On the other hand, as shown in Fig.~\ref{figoe}(b), the RMF(BCS)$^{*}$ approach can reproduce the OES of charge radii of indium isotopes well. RMF(BCS) fails to reproduce that below $N=60$, and even gives an inverted sign around $N=73$. The HF(BCS) model fails to give a reasonable description of these variations.

\begin{figure}[htbp]
\includegraphics[scale=0.6]{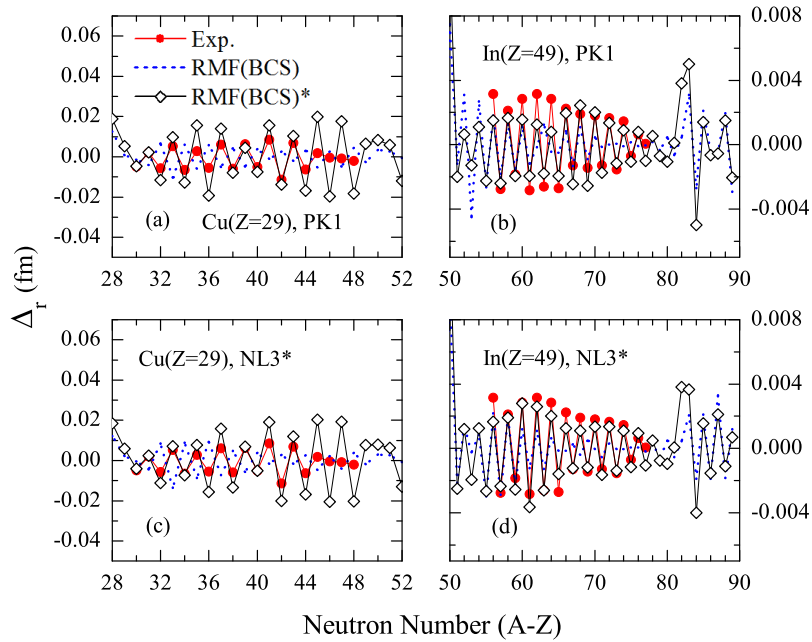}
    \caption{(color online) Odd-even staggering of nuclear charge radii obtained using the PK1~\cite{PhysRevC.69.034319} and NL3$^{*}$~\cite{LALAZISSIS200936} parameter sets for copper (a, c) and indium (b, d) isotopes. The labels are the same as Fig~\ref{figoe}. The experimental data stem from Refs.~\cite{ANGELI201369,deGroote,Sahoo_2020,LI2021101440}.} \label{figoe11}
\end{figure}

In Fig.~\ref{figoe11}, the degree of OES of charge radii for copper and indium isotopes is shown with effective forces PK1 and NL3$^{*}$. One can find that the RMF(BCS) approach with these two parameter sets cannot give the odd-even oscillation behaviors in copper isotopes. Toward the neutron-rich region of indium isotopes, the same scenarios are encountered in both forces. By contrast, the RMF(BCS)$^{*}$ approach with the PK1 and NL3$^{*}$ parameter sets can reproduce the general OES behaviors along copper and indium isotopic chains, but slightly overestimates the amplitudes of OES in copper isotopes. As discussed above, this may result from the $np$-correlations originating from unpaired neutrons and protons.

As discussed above, there is a signal that locates the strong shell effects, namely the abrupt changes of charge radii across the neutron-rich shell closures~\cite{ANGELI201369,An:2020qgp}. This results in the apparently inverted parabolic-like behavior of nuclear charge radii along isotopic chains between two strong neutron closed-shells. The exact description of the nuclear charge radius is still a long-outstanding and open topic in nuclear physics. As one of the important input quantities in astrophysics, reliable predictions of charge radii of finite nuclei play an important role in theoretical studies~\cite{ARNOULD2020103766}. It is therefore necessary to give a unified approach for predicting the nuclear charge radii throughout the whole nuclear chart.

\section{SUMMARY AND OUTLOOK}\label{sec3}
In this work, firstly a modified formula is extended to study charge radii of odd-$Z$ copper and indium isotopic chains. The NL3, PK1 and NL3$^{*}$ parameter sets are employed to investigate the systematic evolution of charge radii with increasing neutron numbers. The RMF(BCS)$^{*}$ approach with these effective forces can reproduce the trend of changes well in copper and indium isotopes. Our results show that the abrupt increases of charge radii occur naturally in copper and indium isotopes beyond $N=50$ and $N=82$, respectively. This conclusion is consistent with the latest data of nuclear charge radii along cadmium and tin isotopic chains, where a discontinuity aspect appears naturally across neutron-rich closed-shells~\cite{Hammen2018,Gorges2019,Goodacre2021}. The inverted parabolic-like behaviors between two neutron shell closures are also reproduced well. The results obtained by these different parameter sets show that our conclusions are almost force-independent.

The modified formula can reproduce the odd-even oscillation behaviors (defined as $\Delta_{r}$), whereas RMF(BCS) and HF(BCS) approaches fail to describe these local variations. As argued in Ref.~\cite{Miller:2018mfb,Ryckebusch2021}, the neutron-proton ($np$) correlations play an important role in determining the fine structure of nuclear charge radii. In our recent works~\cite{An:2020qgp,An:2021rlw}, the modified term associated with Cooper pairs condensation is employed to capture neutron-proton interactions phenomenologically. The odd-even oscillation behaviors, however, are overestimated in copper isotopes, especially close to the $N=50$ shell. This can be attributed to the omission of the $np$-correlations originating from unpaired neutrons and protons in odd-$Z$ isotopes. Meanwhile, one should note that the OES is well reproduced in odd-$Z$ indium isotopes. This indicates that $np$-correlations coming from unpaired neutrons and protons are reduced with the increasing mass. Further study is urgently required.

\section{Acknowledgements}
This work is supported by the Reform and Development Project of Beijing Academy of Science and Technology under Grant No. 13001-2110. This work is also supported in part by the National Natural Science Foundation of China under Grants No. 12135004, No. 11635003, No. 11961141004, No. 12047513. X. J. is grateful for the support of the National Natural Science Foundation of China under Grants No. 11705118. L.-G. C. is grateful for the support of the National Natural Science Foundation of China under Grants No. 11975096 and the Fundamental Research Funds for the Central Universities (2020NTST06).

\bibliography{refs_odd}
\end{document}